\documentclass[conference]{IEEEtran}
\IEEEoverridecommandlockouts
\usepackage{cite}
\usepackage{amsmath,amssymb,amsfonts}
\usepackage{algorithmic}
\usepackage{graphicx}
\usepackage{textcomp}
\usepackage{xcolor}
\usepackage{mathtools}
\usepackage{listings}
\usepackage{hyperref}
\usepackage{hyperref}
\usepackage[normalem]{ulem}
\mathtoolsset{showonlyrefs}
\usepackage{siunitx}

\definecolor{darkbrown}{rgb}{0.36, 0.10, 0.10}

\def\BibTeX{{\rm B\kern-.05em{\sc i\kern-.025em b}\kern-.08em
    T\kern-.1667em\lower.7ex\hbox{E}\kern-.125emX}}

\begin{document}

\title{
On Estimation of Angles of Arrival in Monostatic ISAC Without Instantaneous Transmit CSI
}

\author{
\IEEEauthorblockN{
Ataher Sams$^{1}$, Simone Di Bari$^{1,2}$, Besma Smida$^{1}$, 
Natasha Devroye$^{1}$, Daniela Tuninetti$^{1}$, and Giorgio Taricco$^{2}$
}
\IEEEauthorblockA{
$^{1}$Department of Electrical and Computer Engineering, 
University of Illinois Chicago, Chicago, IL, USA \\
$^{2}$Department of Electronics and Telecommunications (DET), 
Politecnico di Torino, 10129 Turin, Italy \\
\textit{\{asams3, sdibar2, smida, devroye, danielat\}@uic.edu}, 
\textit{gtaricco@ieee.org}
}
}

\maketitle

\begin{abstract}
This paper explores the fundamental limits of Integrated Sensing and Communication (ISAC) in a more realistic setting compared to previous literature when the Base Staion (BS) has only statistical CSI of the communication user rather than full CSI. We analyze a monostatic setting where the BS performs multi-target Angle of Arrival (AoA) estimation while simultaneously communicating with one of the targets. We assume that the BS has statistical CSI about all AoAs, with less uncertainty in the AoA of the communication receiver. The communication receiver is assumed to have perfect CSI. Utilizing a Bayesian Cramér-Rao Bound (BCRB) framework to characterize the fundamental limits of sensing under minimum mean square error (MMSE) criteria, we derive achievable BCRB-rate trade-off regions. Our approach introduces a number of transmission strategies that share power across sensing and communication beams over a coherence time. Our analysis reveals that beam allocation strategies leveraging the principal eigenvectors of the target-specific sensing matrices minimize individual AoA estimation errors, while strategies balancing sensing and communication directions optimize joint estimation performance at the cost of individual accuracy. We demonstrate that leveraging updated BCRB-based sensing information for the communication receiver, due to its lower channel uncertainty, enables significantly improved communication rates.
\end{abstract}


\section{Introduction}
The conceptualization of Integrated Sensing and Communication (ISAC) can be traced back to coexistence and coordination between radar and communication systems within the spectrum. Early research explored the potential synergies between these functionalities, leading to a rapid expansion of work in this field, as highlighted in surveys such as \cite{03_Survey_ISAC, ISAC6G}. 
Here, we focus on the problem of the exploration of optimal trade-offs between sensing and communication capabilities presented in~\cite{ISAC6G,fundamentalPaper, isac_mimo_2, 02_IoT}, where there are some targets to sense and other separate ones to simultaneously communicate with, under various assumption regarding channel state knowledge at various terminals.
In particular, we consider a monostatic scenario where the transmitter is also the sensing receiver that only has statistical knowledge about the targets to estimate.

The work in~\cite{fundamentalPaper}, which is the foundation for our work, addresses the fundamental trade-off in ISAC from both information-theoretic and estimation-theoretic perspectives for vector AWGN channels and the MMSE sensing metric.
The authors in~\cite{fundamentalPaper} define the BCRB-rate region as the set of all possible achievable pairs of ergodic communication rate and sensing MMSE. 
An inner bound can be formed through a time-sharing strategy between two optimal operating points: one where sensing is optimized and one where communication rate is optimized. This simple  baseline strategy already captures the key trade-offs in ISAC. At the sensing-optimized point, the transmit signal should ``align" with the sensing channel, with the input being ``relatively deterministic" to improve sensing accuracy. Conversely, for the rate-optimized point, the transmit signal should ``align" with the communication channel, with the input being ``as random as possible" to maximize communication rate. 
Improved achievable regions were obtained by shaping the input covariance matrix and achieving an improved takeoff compared to time-sharing.

In this paper, we build upon the AoA estimation framework introduced in~\cite{fundamentalPaper}, considering a scenario where the transmitter has access only to statistical Channel State Information (CSI) for both the sensing targets and the communication receivers. Our work is substantially distinguished by the assumption that the full CSI of the communication channel is not known at the transmitting base station, and that the communication users form a subset of the sensing targets. This contrasts with prior literatures like \cite{fundamentalPaper, howmanysensing}, where sensing and communication targets are typically treated as separate entities, allowing the assumption of known CSI for communication user at the transmitter. In our case, the sensing for the Angle of Arrival (AoA) of the targets can be leveraged to increase the communication rate for the communication user. Although the work in \cite{howmanysensing} operates within a comparable context, they focus primarily on optimizing the required number of dedicated sensing beams subject to a communication rate requirement for the user. In our work, the level of channel uncertainty then differs between these two types of targets. Specifically, the uncertainty associated with the communication receivers becomes lower than that of the sensing targets. This difference arises because the transmitter can leverage the sensing (in our case, through Bayesian Cramér-Rao Bound) for the communication target, leading to more precise estimations when transmitting communication data. Communication channels are typically acquired earlier, allowing the transmitter to have a more accurate, though still imperfect, representation of these channels. In contrast, sensing relies on partially outdated CSI, which introduces greater uncertainty. This distinction mirrors practical ISAC scenarios, where a transmitter operates under different levels of CSI accuracy depending on the function being performed. The system must thus account for these varying degrees of uncertainty when optimizing its performance. 

To address this challenge, we propose transmission strategies tailored for the two-target case, which result in achievable ISAC regions, which also include sensing-only, and communication-only scenarios as special cases. Furthermore, we shed light into the trade-offs between sensing and communication performance of the proposed schemes by analyzing how the available power is allocated among the different tasks. By exploring various power allocation through different transmit strategies, we examine how resource distribution impacts the achievable ISAC regions. Our analysis highlights the fundamental trade-offs between sensing accuracy and communication efficiency, demonstrating the crucial role of power allocation and beam design in optimizing overall realistic system performance.

{\bf Paper Organization:}
Section~\ref{sec:model} introduces the system model and problem formulation.  
Section~\ref{sec:BCRB} presents the Bayesian Cramér–Rao Bound framework for the joint estimation of AoAs.  
Section~\ref{sec:outer} derives an ISAC region computable outer bounds.  
Section~\ref{ss: a_s_third} describes the proposed transmission strategy based on optimizing a non-zero-mean Gaussian input and its rationale. 
Section~\ref{sec:Results} provides detailed numerical analysis for our proposed strategy where we  highlight the impact of various system parameters.  
Section~\ref{sec:Conclusion} concludes the paper and discusses directions for future work.

{\bf Notation:}
We denote matrices by bold uppercase letters (e.g., $\mathbf{A}$), vectors by bold lowercase letters (e.g., $\mathbf{a}$), and scalars by regular letters (e.g., $a$). The transpose and Hermitian transpose are written as $(\cdot)^{\rm T}$ and $(\cdot)^{\mathrm{H}}$, respectively. The trace and determinant of a matrix are represented as $\operatorname{Tr}(\cdot)$ and $|\cdot|$, while $\lambda_{\max}(\mathbf{A})$ denotes the maximum eigenvalue of $\mathbf{A}$. The expectation operator is $\mathbb{E}[\cdot]$, while $\Re\{\cdot\}$ and $\Im\{\cdot\}$ indicate the real and imaginary parts of a complex quantity. $\mathcal{CN}(\mathbf{m},\mathbf{R})$ denotes a circularly symmetric proper complex Gaussian vector with mean $\mathbf{m}$ and covariance $\mathbf{R}$. 
All other symbols are defined in the text where first introduced.

\section{System Model}
\label{sec:model}

\begin{figure}
\centering
\includegraphics[width=0.9\linewidth]
{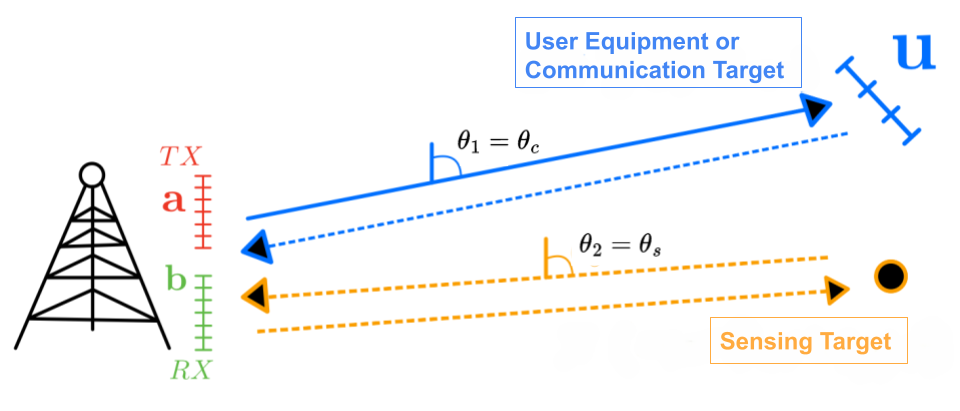}
\caption{Representation of the considered setting: one communication target (UE) that is also a sensing target and a separate sensing-only target.
}
\label{fig:ProbSetting}
\end{figure}

We consider a scenario where a Base Station (BS) is simultaneously a transmitter for the communication receiver and a monostatic radar for the sensing targets,
as shown in Fig.~\ref{fig:ProbSetting} for the case of two targets. 
The BS features a full-duplex uniform linear array (ULA) with $M_\textrm{TX}$ transmitting and $M_\textrm{RX}$ receiving elements, characterized by steering vectors $\mathbf{a} \in \mathbb{C}^{M_\textrm{TX}\times 1}$ and $\mathbf{b} \in \mathbb{C}^{M_\textrm{RX}\times 1}$, respectively. We assume ideal full-duplex operation without self-interference at the BS. 
The communication receiver (or User Equipment, UE) has a ULA with $M_\textrm{UE}$ elements and steering vector $\mathbf{u} \in \mathbb{C}^{M_\textrm{UE}\times 1}$. 
Both the UE and the sensing targets reflect the BS-transmitted signal back to the BS receiver. We assume that the BS has only statistical CSI for both the sensing target and the UE. 
The sensing task is to estimate the AoAs of both of these targets. 
For communication, this sensing information becomes beneficial, as the AoA estimate of the UE obtained through sensing can be used to reduce the channel uncertainty compared to operating without such information.

Let $T$ be the channel coherence time, i.e., the channel parameters undergo synchronous and i.i.d. variations every $T$ channel uses. 
The BS transmits signal $\mathbf{X} \in \mathbb{C}^{M_\textrm{TX}\times T}$, and receives reflected signals $\mathbf{Y}_s \in \mathbb{C}^{M_\textrm{RX}\times T}$ from sensing targets, while the UE receives $\mathbf{Y}_c \in \mathbb{C}^{M_\textrm{UE}\times T}$, where
\begin{align}
    \mathbf{Y}_{c} &= \mathbf{H}_{c} \mathbf{X} + \mathbf{Z}_{c}, \quad 
    \mathbf{H}_{c} = \alpha_1 \ \mathbf{u}(\theta_1) \mathbf{a}^\mathrm{T}(\theta_1), 
\label{eq:Yc} 
    \\
    \mathbf{Y}_s &= \mathbf{H}_s \mathbf{X} + \mathbf{Z}_s, 
    \quad 
    \mathbf{H}_s = \sum_{\ell\in[N_s]} \beta_\ell \ \mathbf{b}(\theta_\ell) \mathbf{a}^\mathrm{T}(\theta_\ell),
\label{eq:Ys}
\end{align}
where $\mathbf{X}$ satisfies the average power constraint
\begin{align}
    \operatorname{Tr}\left\{\mathbb{E}\{\mathbf{R}_{\mathbf{X}}\}\right\} \leq P_{\max}, \quad \mathbf{R}_{\mathbf{X}} := \frac{1}{T} \mathbf{X} \mathbf{X}^\mathrm{H}. 
\label{eq:powconstr}
\end{align}
The noise terms $\mathbf{Z}_c$ and $\mathbf{Z}_s$ are i.i.d., circularly symmetric Gaussian with zero mean and variances $\sigma_c^2$ and $\sigma_s^2$, respectively. The downlink channel matrix $\mathbf{H}_c \in \mathbb{C}^{M_\textrm{UE} \times M_\textrm{TX}}$ represents the communication link, where $\theta_1 \in [0, 2\pi]$ is the AoA of the UE, and $\alpha_1 \in \mathbb{C}$ is the attenuation; $\alpha_1, \theta_1$ are known at the UE. The sensing channel matrix $\mathbf{H}_s \in \mathbb{C}^{M_\textrm{RX} \times M_\textrm{TX}}$ models reflections from $N_s$ targets, with AoAs $\boldsymbol{\theta} = (\theta_1, \theta_2, \ldots, \theta_{N_s})$ and gains $\boldsymbol{\beta} = (\beta_1, \beta_2, \ldots, \beta_{N_s})$. The BS estimates $\boldsymbol{\theta}$, while the attenuation vector $\boldsymbol{\beta}$ is considered as nuisance parameter. We assume $N_s$ known at the BS and the channel parameters to be mutually independent, i.e.,
\begin{align}
    P_{\theta_1, \ldots, \theta_{N_s}, \beta_1, \ldots, \beta_{N_s}} = \prod_{\ell\in[N_s]} P_{\theta_\ell} P_{\Re\{\beta_\ell\}} P_{\Im\{\beta_\ell\}}. 
\label{eq:prior}
\end{align}

\textbf{Sensing Task.} The sensing task consists of estimating the angle of arrivals $\boldsymbol{\theta}$ in~\eqref{eq:Ys}.
As a metric for this sensing task, we use the BCRB~\cite{fundamentalPaper}, a lower bound for the Mean Squared Error (MSE) of weakly unbiased estimators. The BCRB is defined as 
$
    \epsilon := \mathbb{E}_{\mathbf{X}}\big\{ \operatorname{Tr}\big\{ \mathbf{J}_{\boldsymbol \theta|\mathbf X}^{-1}\big\}\big\},
$ 
where $\mathbf{J}_{\boldsymbol \theta|\mathbf X}$ is the Bayesian Fisher Information Matrix (BFIM) of the parameters $\boldsymbol{\theta}$. The conditioning over $\mathbf{X}$ is because, in a monostatic setting, the sensing transmitter is also the sensing receiver so the transmit signal $\mathbf{X}$ is  known. 
The BFIM $\mathbf{J}_{\boldsymbol{\theta} \mid \mathbf{X}}$ of the parameters $\boldsymbol{\theta}$ is given by \cite{BFIM},
\begin{align}
&\mathbf{J}_{\boldsymbol{\theta} \mid \mathbf{X}} := 
\mathbb{E}\left\{\frac{\partial \ln p_{\boldsymbol{\theta}}(\boldsymbol{\theta})}{\partial \boldsymbol{\theta}} \frac{\partial \ln p_{\boldsymbol{\theta}}(\boldsymbol{\theta})}{\partial \boldsymbol{\theta}^{\mathrm{T}}}\right\} + 
\label{eq:BFIM}
\\&
\mathbb{E}\left\{\left.\frac{\partial \ln p_{\mathbf{Y}_{\mathrm{s}} \mid \mathbf{X}, \boldsymbol{\theta}}\left(\mathbf{Y}_{\mathrm{s}} \mid \mathbf{X}, \boldsymbol{\theta}\right)}{\partial \boldsymbol{\theta}} \frac{\partial \ln p_{\mathbf{Y}_{\mathrm{s}} \mid \mathbf{X}, \boldsymbol{\theta}}\left(\mathbf{Y}_{\mathbf{s}} \mid \mathbf{X}, \boldsymbol{\theta}\right)}{\partial \boldsymbol{\theta}^{\mathrm{T}}} \right\rvert\, \mathbf{X}\right\}. \notag
\end{align}
For the AWGN model, $\mathbf{J}_{\boldsymbol{\theta} \mid \mathbf{X}}$ depends on the input $\mathbf{X}$ through the (random in general) sample covariance matrix $\mathbf{R}_{\mathbf{X}}$ in~\eqref{eq:powconstr}.

\textbf{Communication Task.} The communication task consists of reliably transmitting information to the UE with received signal $\mathbf{Y}_c$ in~\eqref{eq:Yc}. As a metric for this task, we use the ergodic achievable rate~\cite{fundamentalPaper}, given by 
$R = \frac{1}{T} I(\mathbf{X}; \mathbf{Y}_{c} | \mathbf{H}_{c})$
where the mutual information is averaged  over the distribution of $\mathbf H_c$, assumed known at the UE, but not the BS. 

\textbf{ISAC region.}
The ISAC region is defined as
\medskip
\begin{align}
\mathcal{R}_{\text{ISAC}} := \bigcup_{\mathcal{X}(P_{\max})} 
\left\{ (\epsilon, R) \; \bigg| \; \epsilon = \mathbb{E}_{\mathbf{X}}\left\{ \operatorname{Tr}\left\{ \mathbf{J}_{\boldsymbol \theta|\mathbf X}^{-1}\right\}\right\}, \right.
\label{eq:epsDef}
\\ 
\left. R = \frac{1}{T} I(\mathbf{X}; \mathbf{Y}_{c} | \mathbf{H}_{c}) \right\},
\label{eq:rateDef}
\end{align} 
where $\mathcal{X}(P_{\max})$ is the set of all possible distributions on $\mathbf{X}$ that satisfy the average power constraint in~\eqref{eq:powconstr}. 
We assume only statistical CSI at the BS transmitter, so $\mathbf{X}$ cannot depend on the realization of $\mathbf{H}_{c}$ or $\mathbf{H}_{s}$ in each coherence time.

\section{Bayesian Cramér-Rao Bound (BCRB)} 
\label{sec:BCRB}
Key for our study is the evaluation the BCRB for the joint estimation of the unknown parameters in the sensing channel matrix: the angles and the complex gains. A general solution to this problem was derived in~\cite{CRB_Optimization}, here we explicitly report the result specialized to our setting.
For $N_s$ sensing targets, each contributing one angle and the real and imaginary parts of its complex gain, the parameter vector is defined as  
\begin{equation}
    \boldsymbol{\eta} = \big[ \boldsymbol{\theta}, \, \Re\{\boldsymbol{\beta}\}, \, \Im\{\boldsymbol{\beta}\} \big] \in \mathbb{R}^{K}, \quad K = 3N_s.
\end{equation}
The corresponding BFIM can be written as~\cite{fundamentalPaper,BFIM}.
\begin{align}
    \mathbf{J}_{{ \boldsymbol \eta} \mid \mathbf X}  
    &= \mathbf{J}^P + \mathbf{F} 
    = 
    \operatorname{Diag} \left[ 
    \mathbf{J}_{\boldsymbol{\theta}}^P,
    \mathbf{J}_{\Re\{\boldsymbol \beta\}}^P,
    \mathbf{J}_{\Im\{\boldsymbol \beta\}}^P
    \right]
    \\& +
    2\begin{bmatrix}
    \Re\left(\mathbf{F}_{11}\right) & \Re\left(\mathbf{F}_{12}\right) & -\Im\left(\mathbf{F}_{11}\right) \\
    \Re\left(\mathbf{F}_{12}\right) & \Re\left(\mathbf{F}_{22}\right) & -\Im\left(\mathbf{F}_{22}\right) \\
   -\Im\left(\mathbf{F}_{11}\right) & -  \Im^\top\left(\mathbf{F}_{22}\right) & \Re\left(\mathbf{F}_{22}\right),
    \end{bmatrix}
\label{eq:for BCRB step 1}
\end{align}
where 
the expressions of $\mathbf{F}_{11}$, $\mathbf{F}_{12}$ and $\mathbf{F}_{22}$  can be found in~\cite[Eq.(14)-(16)]{CRB_Optimization} and are not reported here for sake of space, and where the prior Fisher information matrix $\mathbf{J}^P$ is diagonal because of the assumption in~\eqref{eq:prior}. 

Assuming the complex amplitudes are circularly symmetric, i.e., 
$\mathbb{E}\{{\beta_i}\}=0$, implies $\mathbb E_{\boldsymbol \beta}\{\mathbf F_{12}  \} = 
\mathbf{0}.$
By choosing the phase reference point of the ULAs such that
$
  \mathbf{a}_\ell^{\mathrm H} (\theta) \dot{\mathbf{a}}_\ell (\theta)
= \mathbf{b}_\ell^{\mathrm H} (\theta) \dot{\mathbf{b}}_\ell (\theta)
= 0, \forall \ell\in[N_s]$, 
and by the independence of the complex amplitudes in~\eqref{eq:prior}, by following~\cite{fundamentalPaper,BFIM} we get
\begin{equation}
    \mathbb{E}_{\boldsymbol{\beta}} \left\{ \mathbf{F}_{11} \right\} 
    = \frac{T}{\sigma_{s}^2} 
    \operatorname{Diag} \left[ 
        \mathbb{E} \left\{ |\beta_i|^2 \right\}  \operatorname{Tr} \left[ \overline{\mathbf{M}}_i \, \mathbf{R}_{\mathbf{X}} \right],  \forall i \in [N_s] 
    \right], 
\label{eq:for_BCRB_step3}
\end{equation}
where $\forall i \in [N_s]$ we have
\begin{align}
    \overline{\mathbf{M}}_i 
    &:= \mathbb{E}_{\theta_i} \left\{ 
    \| \dot{\mathbf{b}}(\theta_i) \|^2 \, \mathbf{a}(\theta_i) \mathbf{a}^{\mathrm{H}}(\theta_i)    +\left\| \mathbf{b}(\theta_i) \right\|^2 \, \dot{\mathbf{a}}(\theta_i) \dot{\mathbf{a}}^{\mathrm{H}}(\theta_i)
    \right\},
\label{eq:compact_BCRB_1}
\end{align}
where $\dot{\mathbf{a}}(\theta) := \partial \mathbf{a} / \partial \theta$ and $\dot{\mathbf{b}}(\theta) := \partial \mathbf{b} / \partial \theta$.
We do not report that expression for $\mathbb{E}_{\boldsymbol \beta} \{ \mathbf{F}_{22} \}$ as it will not be needed next. 

Since we are only interested in estimating the angles of arrival, the equivalent BFIM by treating the complex amplitudes as nuisance parameters \cite{fundamentalPaper,BFIM} is given by
\begin{align}
    &{\mathbf{J}^{\rm{(equiv)}}_{\boldsymbol{\theta} \mid \mathbf{X}}}
    = 2 \mathbb{E}_{\boldsymbol \beta}\left\{\mathbf{F}_{11}\right\}
    + \mathbf{J}_{\boldsymbol \theta}^P,
\label{eq:for BCRB step 5}   
\end{align}
which is a diagonal matrix.
Let ${J}_{\theta_i}^P$ be the prior FIM for angle $\theta_i$ and $\epsilon_i(\mathbf{R}_{\mathbf{X}})$ be the BCRB for the $i$-th target, with
\begin{align}
\epsilon_i\left(\mathbf{R}_{\mathbf{X}}\right):=\left(\frac{2 T}{\sigma_s^2} \mathbb{E}\left\{\left|\beta_i\right|^2\right\} 
 \operatorname{Tr} \left[ \overline{\mathbf{M}}_i \, \mathbf{R}_{\mathbf{X}} \right]
+J_{\theta_i}^P\right)^{-1},
\end{align}
then the BCRB for the AoAs is
\begin{align}
\epsilon 
&= \mathbb{E}_{\mathbf{X}}\left\{ 
\operatorname{Tr}\!\left[({\mathbf{J}^{\rm{(equiv)}}_{\boldsymbol{\theta} \mid \mathbf{X}}})^{-1}\right] 
\right\} =  
\mathbb{E}_{\mathbf{X}}\left\{
\sum_{i\in[N_s]}
\epsilon_i(\mathbf{R}_{\mathbf{X}})
\right\}.
\label{eq:BCRB general}
\end{align}

\section{Outer Bound}
\label{sec:outer}
An easily computable outer bound is as follows
\begin{align}
\mathcal{R}_{\text{ISAC}} \subseteq \Big\{ 
(\epsilon, R) \; \big| \; 
\epsilon \geq \sum_{i\in[N_s]} \epsilon_{\min,i}^\prime , 
\quad
R \leq C{^\prime}
\Big\}.
\end{align} 

Here, $\epsilon'_{\min,i}$ is computed as
\begin{align}
\epsilon_{\min,i} 
&:=   \min_{\mathbf R_{\mathbf X}:\ \operatorname{Tr}\left\{\mathbb{E}\left\{\mathbf{R}_{\mathbf{X}}\right\}\right\}
    \leq P_{\max}}
    \mathbb{E}\left\{
    \epsilon_i(\mathbf{R}_{\mathbf{X}})
    \right\}
\label{eq:2nd best BCRB} 
\\&=  \min
  \mathbb{E}_{\mathbf{X}}\left\{
  \frac{1}{\frac{2 T }{\sigma_s^2} \mathbb{E}\left\{|\beta_i|^2\right\} 
{\rm Trace}[\overline{\mathbf{M}}_i \mathbf{R}_{\mathbf{X}} ]
+ J_{\theta_i}^P}
\right\}
  \\&\stackrel{\rm(JI)}{\geq}  
  \min
  \frac{1}
  {\frac{2 T }{\sigma_s^2} \mathbb{E}\left\{|\beta_i|^2\right\} 
  \mathbb{E}_{\mathbf{X}}\left\{
{\rm Trace}[\overline{\mathbf{M}}_i \mathbf{R}_{\mathbf{X}} ]
\right\}
+ J_{\theta_i}^P} 
\label{eg:Jensen_step}
  \\&=  
  \frac{1}{\frac{2 T }{\sigma_s^2} \mathbb{E}\left\{|\beta_i|^2\right\} 
  \max \mathbb{E}_{\mathbf{X}}\left\{
{\rm Trace}[\overline{\mathbf{M}}_i \mathbf{R}_{\mathbf{X}} ]
\right\}
+ J_{\theta_i}^P}
  \\&=  
  \frac{1}{\frac{2 T }{\sigma_s^2} \mathbb{E}\left\{|\beta_i|^2\right\} 
  P_{\max} \lambda_{\max} [\overline{\mathbf{M}}_i]
+ J_{\theta_i}^P}
=: \epsilon_{\min,i}^\prime,
\label{eq:3rd best BCRB}
\end{align}
where the inequality 
marked with `{\rm(JI)}' follows from Jensen’s inequality, 
and the last equality is attained by $\mathbf{X} = \sqrt{P_{\max}} \mathbf{v}_i^{(1)} \otimes [1,1,...,1]$ where $\mathbf{v}_i^{(1)}$ is the eigenvector of $\overline{\mathbf{M}}_i$ corresponding to $\lambda_{\max} [\overline{\mathbf{M}}_i]$, and $[1,1,...,1]$ is the all-one vector of length $T$.
Note, here the (JI)-step is tight.

Finally, $C'$ is computed as
\begin{align} 
    C &:= \max_{\mathbf R_{\mathbf X}:\ \operatorname{Tr}\left\{\mathbb{E}\left\{\mathbf{R}_{\mathbf{X}}\right\}\right\}
    \leq P_{\max}} R, 
    \label{eq:opt_eq}
    \\
    R &:=  
    \mathbb{E}_{\alpha_1, \theta_1} \left\{\log\left(1  + \frac{|\alpha_1|^2}{\sigma^2_c}   \mathbf{a}^{\rm H}(\theta_1) \mathbb{E}\{\mathbf R_{\mathbf X}\} \mathbf{a}(\theta_1)\right) \right\},
    \label{eq:opt_eq2}
    \end{align}
where $C$ is attained by a zero-mean Gaussian input with covariance matrix $\mathbb{E}\{\mathbf R_{\mathbf X}\}$;
$C$ can be upper bounded as follows, for ${\mathbf K}_{\mathbf X} := \mathbb{E}\{\mathbf R_{\mathbf X}\}$ and ${\overline{\mathbf L}_{1} := \mathbb{E}_{\theta_1} [  \mathbf{a}(\theta_1) \mathbf{a}^{\rm H}(\theta_1)] }$:
\begin{align}
C&=\max 
\mathbb{E}_{\alpha_1, \theta_1} \left\{\log\left(1  + \frac{|\alpha_1|^2}{\sigma^2_c}  \mathbf{a}^{\rm H}(\theta_1){\mathbf K}_{\mathbf X} \mathbf{a}(\theta_1)\right) \right\}
\\&\stackrel{\rm(JI)}{\leq}
\max
\log\left(1  + \frac{\mathbb{E}\{|\alpha_1|^2\}}{\sigma^2_c}   \mathbb{E}_{\theta_1} \{ \mathbf{a}^{\rm H}(\theta_1){\mathbf K}_{\mathbf X} \mathbf{a}(\theta_1) \} \right) 
\label{eq:jensen_comm}
\\&=
\log\left(1  + \frac{\mathbb{E}\{|\alpha_1|^2\}}{\sigma^2_c}  \max
{\rm Trace}[
\overline{\mathbf L}_{1} {\mathbf K}_{\mathbf X} 
]
\right) 
\\&=
\log\left(1  + \frac{\mathbb{E}\{|\alpha_1|^2\}}{\sigma^2_c}  
P_{\max} \lambda_{\max} [\overline{{\mathbf L}}_{1}]
\right) =: C^\prime,
\label{eq:c_prime}
\end{align}
where the last equality is attained by
$\mathbf{X} = \sqrt{P_{\max}} \mathbf{l}_{1} \otimes [G_1,G_2,...,G_T]$ where $\mathbf{l}_{1}$ is the eigenvector of $\overline{\mathbf{L}}_1$ corresponding to $\lambda_{\max} [\overline{\mathbf{L}}_1]$, and $[G_1,G_2,...,G_T]$ has $T$ i.i.d. $\mathcal{CN}(0,1)$ components.
Note, here the (JI)-step is not tight.

\section{Achievable Strategy} 
\label{ss: a_s_third}
We consider a transmit signal $\mathbf{X}$ that should have two end goals: achieving low estimation error on the AoAs and ensuring a high communication rate. In an attempt to span between the (deterministic) sensing-optimal strategy and the (Gaussian) communication-optimal strategy, in this work we consider optimizing over the non-zero-mean Gaussian input
\begin{align} 
    \mathbf{X} = \left[\mathbf{x}_1, \ldots, \mathbf{x}_T\right]: \mathbf{x}_t 
    \sim \mathcal{CN}\left( 
    \sqrt{P_{s,t}}\, \mathbf{s}_t, \
    P_{c,t} \mathbf{c}_t  \mathbf{c}_t^{\rm H}
    \right),
\label{eq:transmit_main}
\end{align}
where \(\mathbf{s}_t\) is the unit-length sensing vector at time \(t\), while \(\mathbf{c}_t\) is the unit-length communication vector at time \(t\) (modulated by i.i.d. \(G_t \sim \mathcal{CN}(0,1)\)) and is subject to the power constraint
\begin{align}
    \frac{1}{T} \sum_{t\in[T]} (P_{s,t} + P_{c,t}) \leq P_{\max}.
\label{eq:power_main}
\end{align}
The choice in~\eqref{eq:transmit_main} let us combine a `deterministic beam' ($\sqrt{P_{s,t}}\,\mathbf{s}_t$) for sensing with a `Gaussian beam' ($\sqrt{P_{c,t}}\,\mathbf{c}_t G_t$) for communication, giving a flexible way to study 
both the subspace and the deterministic-vs-random traedoffs in ISAC.

Thus, for an achievable ISAC region with the transmitted signal in~\eqref{eq:transmit_main},
we compute the ergodic achievable rate $R$ as
\begin{align}
R
&= \frac{1}{T} \sum_{t=1}^{T} \mathbb{E}_{\theta_1,\alpha_1} \!\!\left\{ \log \left( 1 + 
\frac{P_{c,t} |\alpha_1|^2}{\sigma_c^2}  \left| \mathbf{a}^{\rm H} (\theta_1) \mathbf{c}_t \right|^2
\right) \!\!\right\},
\label{eq:communication_eq}
\end{align}
and the BCRB as $\epsilon= \sum_{i\in[N_s]} \overline{\epsilon}_{i}$, where
\begin{align} 
    &\overline{\epsilon}_{i}
    :=\mathbb{E}_{\mathbf{X}}\left\{\left(\frac{2 T}{\sigma_{s}^2}  
    \mathbb{E}\left\{\left|\beta_i\right|^2\right\}
    f_i(\mathbf{R}_{\mathbf{X}}) 
    +J_{\theta_i}^{\mathrm{P}}\right)^{-1}\right\},
\label{BCRB_main_Equ}
\\
&f_i(\mathbf{R}_{\mathbf{X}})
:=
\frac{1}{T}\sum_{t\in[T]} 
(\sqrt{P_{s,t}}\, \mathbf{s}_t + \sqrt{P_{c,t}}\, \mathbf{c}_t G_t)^{\rm H}
\\&\qquad\qquad
\overline{\mathbf{M}}_i
(\sqrt{P_{s,t}}\, \mathbf{s}_t + \sqrt{P_{c,t}}\, \mathbf{c}_t G_t) 
=\operatorname{Tr}[\overline{\mathbf{M}}_i \mathbf{R}_{\mathbf{X}}].
\label{Rx_BCRB_main_Equ}
\end{align} 
for $\overline{\mathbf{M}}_i$ defined in~\eqref{eq:compact_BCRB_1}, 
and $\mathbf{R}_{\mathbf{X}}$ is evaluated from~\eqref{eq:transmit_main}
with i.i.d. \(G_t \sim \mathcal{CN}(0,1)\).
We next examine two special cases.

{\bf Special Case 1: deterministic only.} 
Here we transmit only deterministic signals ($P_{c,t} = 0$).
This strategy provides insights into minimum bounds on estimation accuracy for joint AoA estimation. For this case, $R=0$ (i.e., without randomness the rate is zero) and $\epsilon = \sum_{i\in[N_s]} \overline{\epsilon}_{\text{d},i}$ where
\begin{align}
    \overline{\epsilon}_{\text{d},i}
&:=
    \left( 
    \frac{2\mathbb{E}\left\{ |\beta_i|^2 \right\}}{\sigma_s^{2}}
    \sum_{t\in[T]} P_{s,t} 
    \mathbf{s}_t^{\rm H}
    \overline{\mathbf{M}}_i 
    \mathbf{s}_t   
    +J_{\theta_i}^{\mathrm{P}}
    \right)^{-1}.
    \label{eq:sensing_eq}
\end{align}

{\bf Special Case 2: Gaussian only.} 
Here no power is allocated to deterministic signals ($P_{s,t} = 0$). 
This setup corresponds to a typical communication-only signal that may also be repurposed for ISAC for both communication and sensing. 
The rate is computed as in~\eqref{eq:communication_eq} and $\epsilon = \sum_{i\in[N_s]} \overline{\epsilon}_{\text{g},i}$, where
\begin{align} 
    \overline{\epsilon}_{\text{g},i}
    &:=
    \mathbb{E}_{\mathbf{X}}\left\{\left(
    \frac{2\mathbb{E}\left\{ |\beta_i|^2 \right\}}{\sigma_{s}^2}  
    \sum_{t\in[T]} P_{c, t} 
    \mathbf{c}_t^{\mathrm{H}}
    \overline{\mathbf{M}}_i
    \mathbf{c}_t \left|G_t\right|^2
    +J_{\theta_i}^{\mathrm{P}}
    \right)^{-1}\right\}.
\end{align} 
No closed form expression exists for $\overline{\epsilon}_{\text{g},i}$ when the directions $\mathbf{c}_1,\ldots,\mathbf{c}_T$  or the powers $P_{c, 1} ,\ldots,P_{c, T}$ are different.



In principle, we aim to evaluate the achievable ISAC region by considering all possible sensing and communication directions in~\eqref{eq:transmit_main}, as well as all power allocations satisfying~\eqref{eq:power_main}. However, this constitutes an extremely large optimization space. We therefore describe next our rationale for choosing certain sensing and communication directions.
We concentrate on the two `corner points' on the ISAC region: one that is optimized for sensing only (Special Case 1) and another that is optimized for communication only (Special Case 2).

{\bf Choice of Sensing Directions.}
We begin by highlighting the trade-offs between estimating a single target and estimating multiple targets jointly.
For a single target, minimizing its BCRB in~\eqref{eq:2nd best BCRB} is equivalent to transmitting along the principal eigenvector $\mathbf{v}_1^{(i)}$ of its sensing information matrix $\overline{\mathbf{M}}_i$ in~\eqref{eq:compact_BCRB_1}, which achieves $\epsilon_{\min,i}^\prime$ in~\eqref{eq:3rd best BCRB}; this captures the direction providing the maximum information about the $i$-th angle. This motivates the use of $\{ \mathbf{v}_1^{(i)}, \cdots, \mathbf{v}_{M_{\rm TX}}^{(i)}, \ i\in[N_s] \}$ as good candidates for `single-angle sensing' directions.
However,  when jointly estimating multiple AoAs, an analytical solution for the minimizer of~\eqref{eq:BCRB general} does not appear possible.
Instead, we numerically 
solve the joint BCRB minimization problem by using CVX, and
indicate the eigenvectors of the optimal (deterministic) sample covariance matrix as $\hat{\mathbf{r}}^{(\rm s)}_1, \cdots, \hat{\mathbf{r}}^{(\rm s)}_{M_{\rm TX}}$, which are good candidates for `joint sensing' directions.

Inspired by~\cite[Corollary 2]{fundamentalPaper}, which claims that under certain conditions the optimal sample covariance matrix for the BCRB minimization problem has rank $\min\{M_{\rm TX},N_s\}$, we construct the `beam' directions
as linear combinations of 
$\{ \hat{\mathbf{r}}^{(\rm s)}_j, 
\mathbf{v}_j^{(i)} : 
\ j\in[\min\{M_{\rm TX},N_s\}], 
\ i\in[N_s] \}$,
where here we assume that the eigenvalues are ordered from largest to smallest, i.e., $\mathbf{v}_1^{(i)}$ is the principal eigenvector of $\overline{\mathbf{M}}_i$, etc.

{\bf Choice of Communication Directions.}
When focusing solely on the communication task, we use CVX to solve the rate maximization problem~\eqref{eq:opt_eq}. We let $\hat{\mathbf{r}}^{(\rm c)}_1,\ldots,\hat{\mathbf{r}}^{(\rm c)}_{M_{\rm TX}}$ be the eigenvectors of the numerically determined optimal covariance matrix. 
Inspired by the notion of degrees of freedom (DoF) in wireless communications \cite{tse2005}, we construct the `beam' directions
as linear combinations of
$\{ \hat{\mathbf{r}}^{(\rm c)}_j : 
\ j\in[\min\{M_{\rm TX},D\}], 
\}$, where $D$ is the DoF of the communication channel, and where here we assume that the eigenvalues are ordered from largest to smallest.

{\bf Example.}
For the case of $N_s=2$ targets, $D=1$ (i.e., here the communication channel is rank~1), and $M_{\rm TX} \geq 2$, the set of candiadte directions used in our design are $\mathbf{v}_1^{(1)}, \mathbf{v}_2^{(1)}, \mathbf{v}_1^{(2)}, \mathbf{v}_2^{(2)}, \hat{\mathbf{r}}^{(\rm s)}_1, \hat{\mathbf{r}}^{(\rm s)}_2$, and $\hat{\mathbf{r}}^{(\rm c)}_1$, as listed in Table~\ref{tab:vector_definitions}. 

\begin{table}
\centering
\caption{Definition and Computation of Direction Vectors}
\label{tab:vector_definitions}
\begin{tabular}{|l|p{5.4cm}|}
\hline
\textbf{Vector(s)} & \textbf{Definition and Computation} \\
\hline
$\mathbf{v}_1^{(1)}, \mathbf{v}_2^{(1)}$ & Principal eigenvectors of $\overline{\mathbf{M}}_1$ via eigen-decomposition; represent dominant sensing directions for target 1. \\
\hline
$\mathbf{v}_1^{(2)}, \mathbf{v}_2^{(2)}$ & Principal eigenvectors of $\overline{\mathbf{M}}_2$ via eigen-decomposition; represent dominant sensing directions for target 2. \\
\hline
$\hat{\mathbf{r}}^{(\rm s)}_1, \hat{\mathbf{r}}^{(\rm s)}_2$ & Principal eigenvectors of optimal $\mathbf{R}_{\mathbf{X}}$ from numerical minimization of~\eqref{eq:BCRB general}; represent optimal joint sensing directions. \\
\hline
$\hat{\mathbf{r}}^{(\rm c)}_1$ & Principle eigenvector of optimal covariance matrix from numerical solution of~\eqref{eq:opt_eq}; represent optimal direction for communication \\
\hline
\end{tabular}
\end{table}

\section{Numerical Analysis}
\label{sec:Results}

In this section, we evaluate the BCRB in~\eqref{BCRB_main_Equ} and the communication rate in~\eqref{eq:communication_eq} for our defined strategy in~in~\eqref{eq:transmit_main}. These evaluations are performed under the following assumptions. We consider a system with \( M_{\text{Tx}} = M_{\text{Rx}} = 10 \) antennas, spaced at half-wavelength intervals. The maximum SNR per antenna is set to 10~dB for sensing and 15~dB for communication. The coherence time is \( T = 2 \). For notational simplicity, we define the AoA of the communication user as \( \theta_1 = \theta_c \), and the AoA of the sensing target as \( \theta_2 = \theta_s \), as illustrated in Fig.~\ref{fig:ProbSetting}. 
Numerical averages are performed using 10,000 realizations of sensing angles, 1,000 realizations of communication angles, and 10,000 Gaussian samples.

AoAs are assumed to follow a Von Mises prior distribution.
This distribution has two parameters $(\bar{\theta}, \kappa)$.
$\kappa>0$ controls the concentration of the distribution around the mean direction $\bar{\theta}$ (i.e., a larger $\kappa$ indicates tighter concentration or smaller uncertainty, while a smaller $\kappa$ corresponds to greater spread). 
In fact, as $\kappa$ increases, the distribution approaches a Gaussian distribution with mean $\bar{\theta}$ and variance $1/\kappa$.
For this reason, in the this work we approximate the (regular) variance of the Von Mises distribution as $\sigma^2 \approx 1/\kappa$.

Other quantities of interest for the BCRB are the prior Fisher information, which can be worked out to be
\begin{equation}
J_\theta^P = \frac{\kappa^2}{2} \left(1 - \frac{I_{2}(\kappa)}{I_{0}(\kappa)} \right),
\end{equation}
where $I_n(\kappa)$ the modified Bessel function of order $n$. 
Unfortunately, no closed-form expression exists for $\overline{\mathbf{M}}_i$ in~\eqref{eq:compact_BCRB_1}.

In this work, we aim to capture the fact that communication targets have already been acquired before the BS begins communicating with them. Consequently, the uncertainty around the mean AoA $\bar{\theta}$ should be smaller than the uncertainty present when that angle was first acquired and estimated. We therefore model the variance of the sensing angle as pre-acquisition and that of the communication angle as post-acquisition.
We express this relationship in terms of the concentration parameters, leading to $\kappa_{c} > \kappa_{s}$.
In particular, we set
\begin{equation}
\kappa_{\rm pre} 
= \kappa_s 
 \leq
\kappa_c
= \kappa_{\rm post}
 \approx
 \frac{1}{ {\rm MMSE}(\theta|\mathbf{Y}_s) },
\end{equation}
where the MMSE for the AoA $\theta$ is computed from the sensing channel output $\mathbf{Y}_s$ assuming a dispersion $\kappa_{\rm pre}$ for $\theta$.
With this, in our numerical evaluations, we set $\bar{\theta}_s=30^{\circ}$, $\bar{\theta}_c=100^{\circ}$, and we set 
$\kappa_{\rm pre}  = \kappa_s \approx 2.184$ leading to 
$\kappa_{\rm post} = \kappa_c \approx 256.674.$ 
This correspond to a pre-acquisition standard deviation of $\approx 38.77^{\circ}$ and
a post-acquisition standard deviation of $\approx 3.58^{\circ}$, which highlights that the post-acquisition communication channel is less uncertain than the pre-acquisition sensing one.

{\bf Power-Allocation Sweep for ISAC Trade-offs.}
To see the trade-offs, we parameterize the transmit design by power allocation parameters $\lambda_i$'s, which are non-negative and constrained as $\sum_i \lambda_i \leq 1, \lambda_i \in [0,1]$, where each $\lambda_i$ represents the the fraction of total power assigned to a specific direction in 
$\{\hat{\mathbf{r}}^{(\rm s)}_1,\hat{\mathbf{r}}^{(\rm s)}_2,\mathbf v_1^{(1)},\mathbf v_2^{(1)}, \mathbf v_1^{(2)},\mathbf v_2^{(2)},\hat{\mathbf{r}}^{(\rm c)}_1\}$ across each channel use within the coherence interval (here $T=2$). So for $L$ possible candidate of transmit directions, we compare operating points at the same total power by restricting to the simplex
\begin{equation}
\Delta_{L-1}\triangleq\Big\{\boldsymbol{\lambda}\in\mathbb{R}_+^{L}:\ \sum_{\ell=1}^{L}\lambda_\ell=1\Big\}.
\end{equation}

We sweep $\boldsymbol{\lambda}$ to trace the ISAC trade–off. For each $\boldsymbol{\lambda}$, we form the beams, build the average transmit covariance $\mathbf{K}_{\mathbf X}=\mathbb{E}\{\mathbf{R}_{\mathbf X}\}$ under the transmit signal in \eqref{eq:transmit_main}, evaluate $R(\boldsymbol{\lambda})$ via the rate expression, and compute the per–angle BCRBs to obtain $\mathrm{CRB}_{\theta_s}(\boldsymbol{\lambda})$, $\mathrm{CRB}_{\theta_c}(\boldsymbol{\lambda})$, and $\sum_i \mathrm{CRB}_{\theta_i}(\boldsymbol{\lambda})$. 
By varying $\boldsymbol{\lambda}$, we allocate power across the candidate directions in Table~\ref{tab:vector_definitions}, which in turn moves the operating point across two 
planes: 
\\(i) the $\mathrm{CRB}_{\theta_s}$–$\mathrm{CRB}_{\theta_c}$ plane, where each point is color-coded by the corresponding rate for the UE, thus showing the joint relationship between sensing accuracy and communication performance, and  
\\(ii) the $R$ vs $\sum_i \mathrm{CRB}_{\theta_i}$ plane, which directly illustrates the trade–off between rate and the aggregate sensing error bound.  
\\These reveal the Pareto frontier of our design, where we can clearly see the trade-off between sensing and communication.

\textbf{Both Deterministic and Gaussian.} 
For the two-target scenario, the sensing optimal transmit covariance matrix has rank two \cite[Corollary 2]{fundamentalPaper}. We define the beamforming directions $\mathbf{s}_1, \mathbf{c}_1$ and $\mathbf{s}_2, \mathbf{c}_2$ as the sensing and communication directions used in the first and second halves of the coherence time, respectively. This structure allows us to transmit in one direction during the first half and another during the second for $T=2$, capturing a wide range of directional approaches within the general strategy in \eqref{eq:transmit_main}, i.e.,
\begin{align}
\mathbf{s}_1&= \sqrt{\lambda_1}\hat{\mathbf{r}}^{\rm(s)}_1 
             + \sqrt{\lambda_2}\hat{\mathbf{r}}^{\rm(s)}_2 
             + \sqrt{\lambda_3}\mathbf{v}_1^{(1)} 
             + \sqrt{\lambda_4}\mathbf{v}_1^{(2)}, \\
\mathbf{c}_1 &= \sqrt{\lambda_5}\hat{\mathbf{r}}^{\rm(c)}_1. 
\label{eq:first_half}
\end{align}
In the second half, the system reuses the same directions with possibly different power allocation:
\begin{align}
\mathbf{s}_2 &= \sqrt{\lambda_6}\hat{\mathbf{r}}^{\rm(s)}_1 
             + \sqrt{\lambda_7}\hat{\mathbf{r}}^{\rm(s)}_2 
             + \sqrt{\lambda_8}\mathbf{v}_1^{(1)} 
             + \sqrt{\lambda_9}\mathbf{v}_1^{(2)}, \\
\mathbf{c}_2 &= \sqrt{\lambda_{10}}\hat{\mathbf{r}}^{\rm(c)}_1.
\label{eq:second_half}
\end{align}
%
We analyze the impact of different spanning choices by selectively nullifying vectors through their corresponding $\lambda$ values (G for General):

\begin{itemize}
    \item Choice~G1: Complete span with all vectors.
    \item Choice~G2: Exclude communication-optimal vectors [\(\lambda_5\!=\!\lambda_{10}\!=\!0\)].
    \item Choice~G3: Exclude principal eigenvectors of \(\overline{\mathbf{M}}_i\) [\(\lambda_3\!=\!\lambda_4\!=\!\lambda_8\!=\!\lambda_9\!=\!0\)].
    \item Choice~G4: Exclude optimal sensing vectors [\(\lambda_1\!=\!\lambda_2\!=\!\lambda_6\!=\!\lambda_7\!=\!0\)].
\end{itemize}

\begin{figure}
    \centering
    \includegraphics[width=\linewidth] {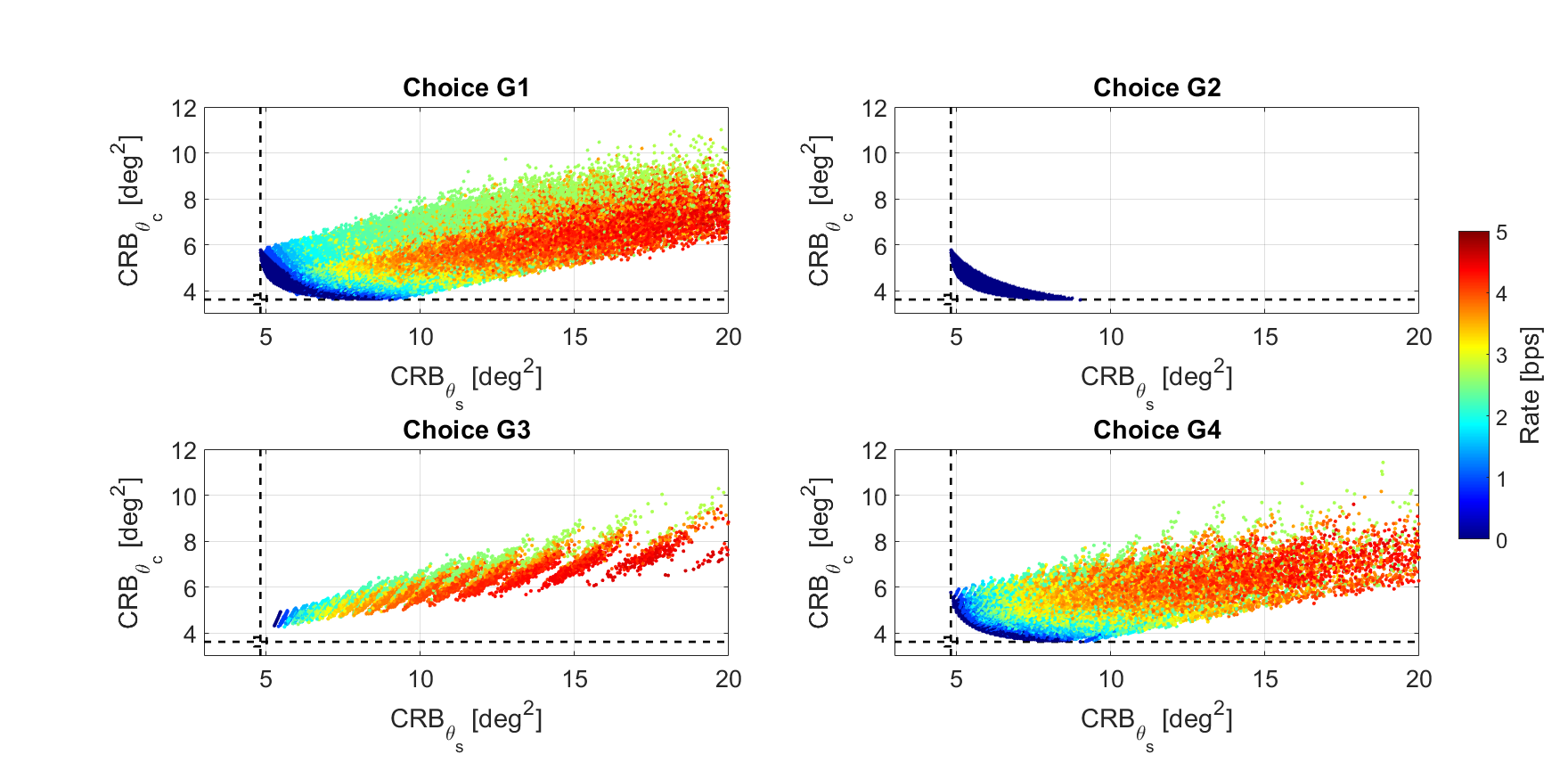}
    \caption{Achievable BCRB-rate regions under Choices G1–G4 for sensing and communication angles $\bar{\theta}_s = 30^\circ$, $\bar{\theta}_c = 100^\circ$, with initial standard deviation $\sigma_{\theta} = 30^\circ$ for both angle. The black lines indicate the minimum achievable BCRB for each target.}
    \label{fig:combined_third}
\end{figure}

For the ISAC scenario analysis, our results as depicted in Fig.~\ref{fig:combined_third} demonstrate several key findings across the choices above. Choice G2, which excludes information-carrying beams, achieves the minimum BCRB. Both Choice G2 and Choice G4 span along the principal eigenvectors of $\overline{\mathbf{M}}_1$ and $\overline{\mathbf{M}}_2$, resulting in minimal estimation error for the individual parameters $\theta_s$ (for sensing target) and $\theta_c$ (for communication user). When prioritizing communication while preserving single-angle estimation accuracy, Choice G4 proves effective for ISAC. Conversely, Choice G3 spans optimal sensing and communication directions, reducing joint angle estimation error at the cost of individual accuracy. Choice G1 captures the complete solution space, showing how communication rate scales with power allocation between sensing and communication vectors. This is consistent with our discussion presented in previous Subsection, where the contribution of each beam to the overall trade-off is explained.

\textbf{Only Deterministic.} We span over the previously mentioned directions and principal eigenvectors for sensing, as well as include the second largest eigenvectors of $\overline{\mathbf{M}}_1$ and $\overline{\mathbf{M}}_2$ to investigate their potential influence on system performance. We now define $\mathbf{s}_1$ and $\mathbf{s}_2$ as,
\begin{align}
\mathbf{s}_1 &= \sqrt{\lambda_1} \hat{\mathbf{r}}^{\rm(s)}_1 
            + \sqrt{\lambda_2} \hat{\mathbf{r}}^{\rm(s)}_2 
            + \sqrt{\lambda_3} \mathbf{v}_1^{(1)} 
            + \sqrt{\lambda_4} \mathbf{v}_2^{(1)} \nonumber \\
            &\quad + \sqrt{\lambda_5} \mathbf{v}_1^{(2)} 
            + \sqrt{\lambda_6} \mathbf{v}_2^{(2)}, \\
\mathbf{s}_2 &= \sqrt{\lambda_7} \hat{\mathbf{r}}^{\rm(s)}_1 
            + \sqrt{\lambda_8} \hat{\mathbf{r}}^{\rm(s)}_2 
            + \sqrt{\lambda_9} \mathbf{v}_1^{(1)} 
            + \sqrt{\lambda_{10}} \mathbf{v}_2^{(1)} \nonumber \\
            &\quad + \sqrt{\lambda_{11}} \mathbf{v}_1^{(2)} 
            + \sqrt{\lambda_{12}} \mathbf{v}_2^{(2)}.
\end{align}
To evaluate the impact of different spanning choices, we examine scenarios where specific vectors are nullified by choosing (S for Sensing):
\begin{itemize}
    \item Choice S1: Complete span with all vectors
    \item Choice S2: Exclude optimal sensing vectors [\(\lambda_1\!=\!\lambda_2\!=\!\lambda_7\!=\!\lambda_8\!=\!0\)]
    \item Choice S3: Exclude all principal eigenvectors of $\overline{\mathbf{M}}_i$ [\(\lambda_3\!=\!\lambda_4\!=\!\lambda_5\!=\!\lambda_6\!=\!\lambda_9\!=\!\lambda_{10}\!=\!\lambda_{11}\!=\!\lambda_{12}\!=\!0\)]
    \item Choice S4: Exclude second main eigenvectors of $\overline{\mathbf{M}}_i$ [\(\lambda_4\!=\!\lambda_6\!=\!\lambda_{10}\!=\!\lambda_{12}\!=\!0\)]
    \item Choice S5: Exclude sensing vectors aligned with the communication target in the first half and with the sensing target in the second half [\(\lambda_3\!=\!\lambda_4\!=\!\lambda_{11}\!=\!\lambda_{12}\!=\!0\)]
    \item Choice S6: Rank-1 beam configuration with the same signals across time slots [\(\lambda_1\!=\!\lambda_7,\lambda_2\!=\!\lambda_8,\lambda_3\!=\!\lambda_9,\lambda_4\!=\!\lambda_{10},\lambda_5\!=\!\lambda_{11},\lambda_6\!=\!\lambda_{12}\)]
\end{itemize}

\begin{figure}
    \centering
    \includegraphics[width=\linewidth]{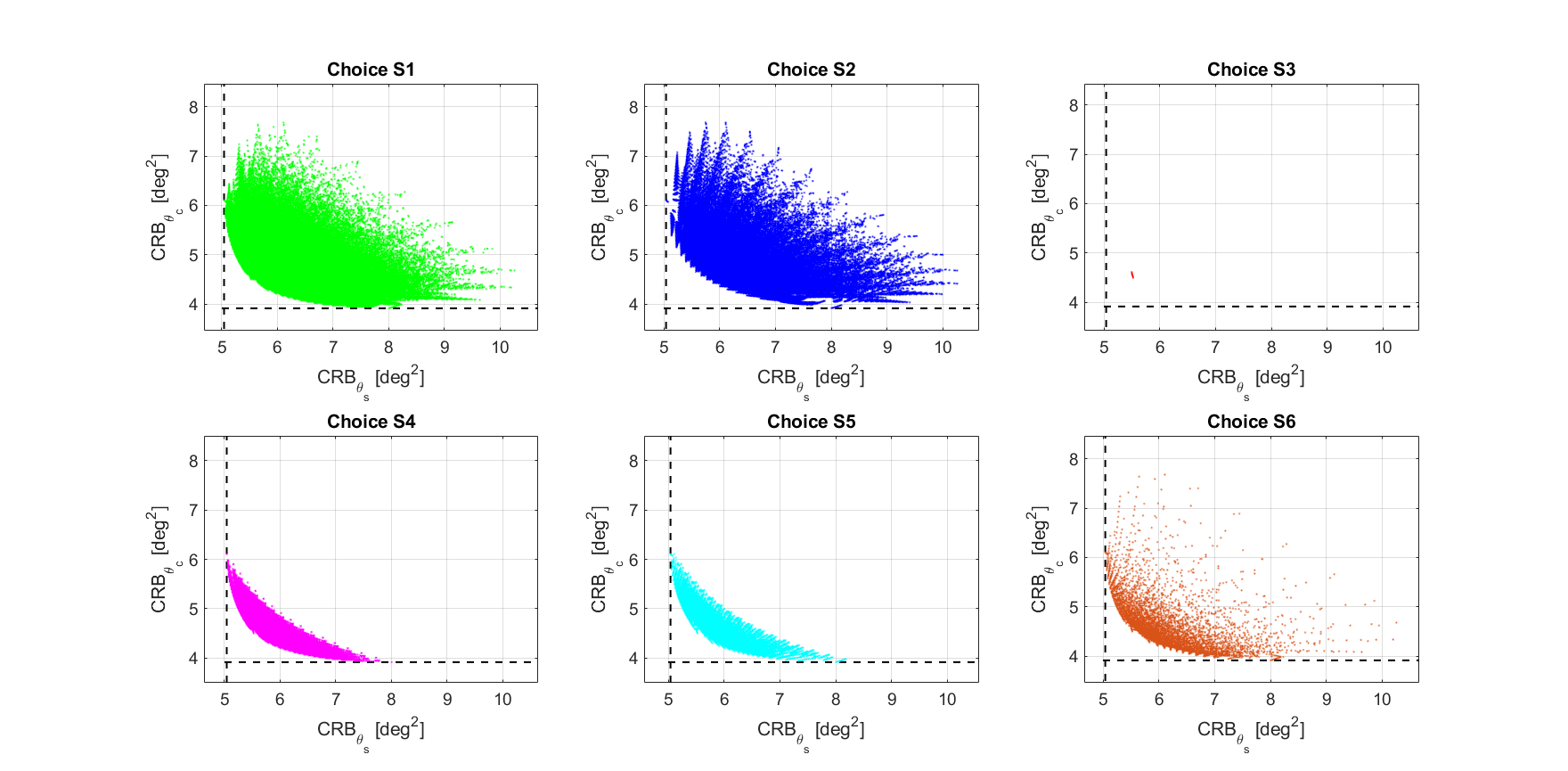}
    \caption{BCRB-rate regions for deterministic or information-less beam scenario under Choices S1–S6 (Special Case 1).}
    \label{fig:ats_final_subplot_sec_10}
\end{figure}
Fig.~\ref{fig:ats_final_subplot_sec_10} reveals how the sensing region evolves across different vector configurations. For this case with only deterministic or information-less signals, we focus exclusively on sensing performance as here rate $R=0$. This highlights the role of the deterministic component of the transmit signal, showing how it directly influences sensing performance. Configurations that utilize principal eigenvectors of $\mathbf{M}_i$ (Choices S2, S4, S5, and S6)- attain optimal estimation performance for individual targets. However, to optimize joint estimation performance, corresponding to the lower-left corner of the BCRB region, it is essential to span across both optimal sensing directions, represented by vectors \( \hat{\mathbf{r}}^{\rm(s)}_1 \) and \( \hat{\mathbf{r}}^{\rm(s)}_2 \), as done in Choices S3, S4, and S6. However, achieving the two individual minimum CRB points simultaneously remains infeasible, highlighting a fundamental trade-off in joint estimation. We also observe that the rank-2 configuration (Choice S1) yields a richer set of solution points in the achievable region compared to the rank-1 configuration (Choice S6). 

\textbf{Only Gaussian.} Although the transmitted signal is purely Gaussian, our goal is to support both communication and sensing. Similar to the previous two scenarios, we include all candidate directions in the linear combinations, since sweeping over the full set produces a broader region of operating points. This will allow us to capture the complete performance envelope. Therefore, the beam directions are constructed as linear combinations of vectors optimized for both sensing and communication:
\begin{equation}
\mathbf{c}_1 = \sqrt{\lambda_1}\hat{\mathbf{r}}^{\rm(s)}_1
            + \sqrt{\lambda_2}\hat{\mathbf{r}}^{\rm(s)}_2
            + \sqrt{\lambda_3}\mathbf{v}_1^{(1)}
            + \sqrt{\lambda_4}\mathbf{v}_2^{(1)}
            + \sqrt{\lambda_5}\hat{\mathbf{r}}^{\rm(c)}_1
\label{eq:c1}
\end{equation}
\begin{equation}
\mathbf{c}_2 = \sqrt{\lambda_6}\hat{\mathbf{r}}^{\rm(s)}_1
            + \sqrt{\lambda_7}\hat{\mathbf{r}}^{\rm(s)}_2
            + \sqrt{\lambda_8}\mathbf{v}_1^{(1)}
            + \sqrt{\lambda_9}\mathbf{v}_2^{(1)}
            + \sqrt{\lambda_{10}}\hat{\mathbf{r}}^{\rm(c)}_1
\label{eq:c2}
\end{equation}

\begin{figure}
    \centering
    \includegraphics[width=\linewidth]{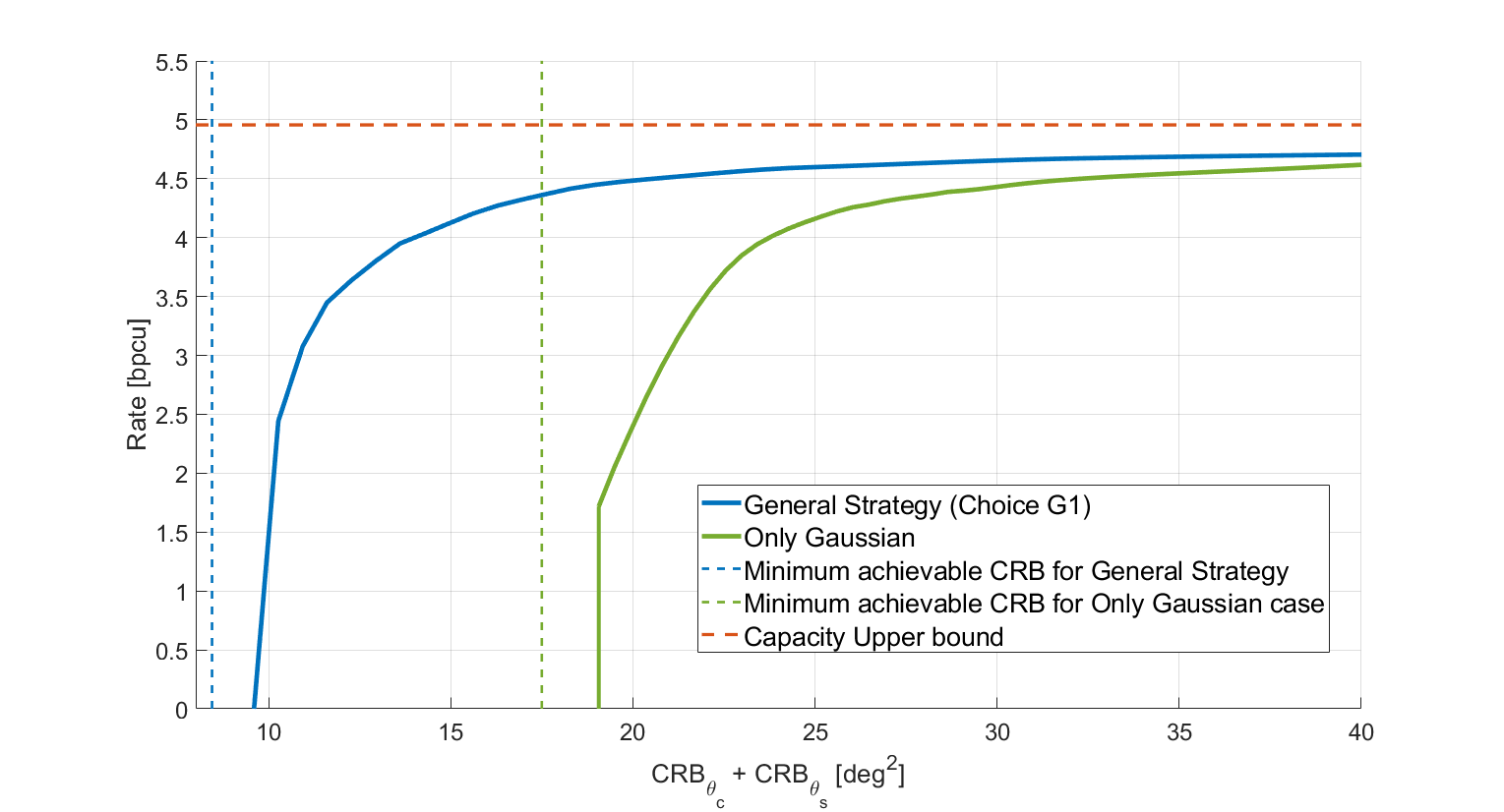}
    \caption{Achievable BCRB-rate regions comparison for main strategy where we transmit both deterministic and Gaussian (Main Strategy in~\eqref{eq:transmit_main}), and only Gaussian. The dashed blue and green lines indicate the minimum achievable CRBs for each case, obtained by computing the sum of individual minimum possible CRBs of both targets. The horizontal red dashed line represents the capacity upper bound $C'$ derived in \eqref{eq:c_prime}. }
    \label{fig:rateVScrb_30_final} 
\end{figure}

 In Fig.~\ref{fig:rateVScrb_30_final}, we present the inner bounds of the BCRB-rate region under the choice of \eqref{eq:transmit_main} and Only Gaussian case, spanning over all possible vectors. Utilizing the updated sensing BCRB for the communication target as described previously in this section to reflect the sensing uncertainty's BCRB when taking the expectation enables substantially higher communication rates. Transmitting only information-carrying signals (special case 2) degrades sensing performance, leading to lower communication rates. This effect is more visible for realistic (lower) sensing SNRs. Again achieving the theoretical optimal performance (combined single best estimation) remains unattainable. 


\begin{figure}
    \centering
    \includegraphics[width=\linewidth]{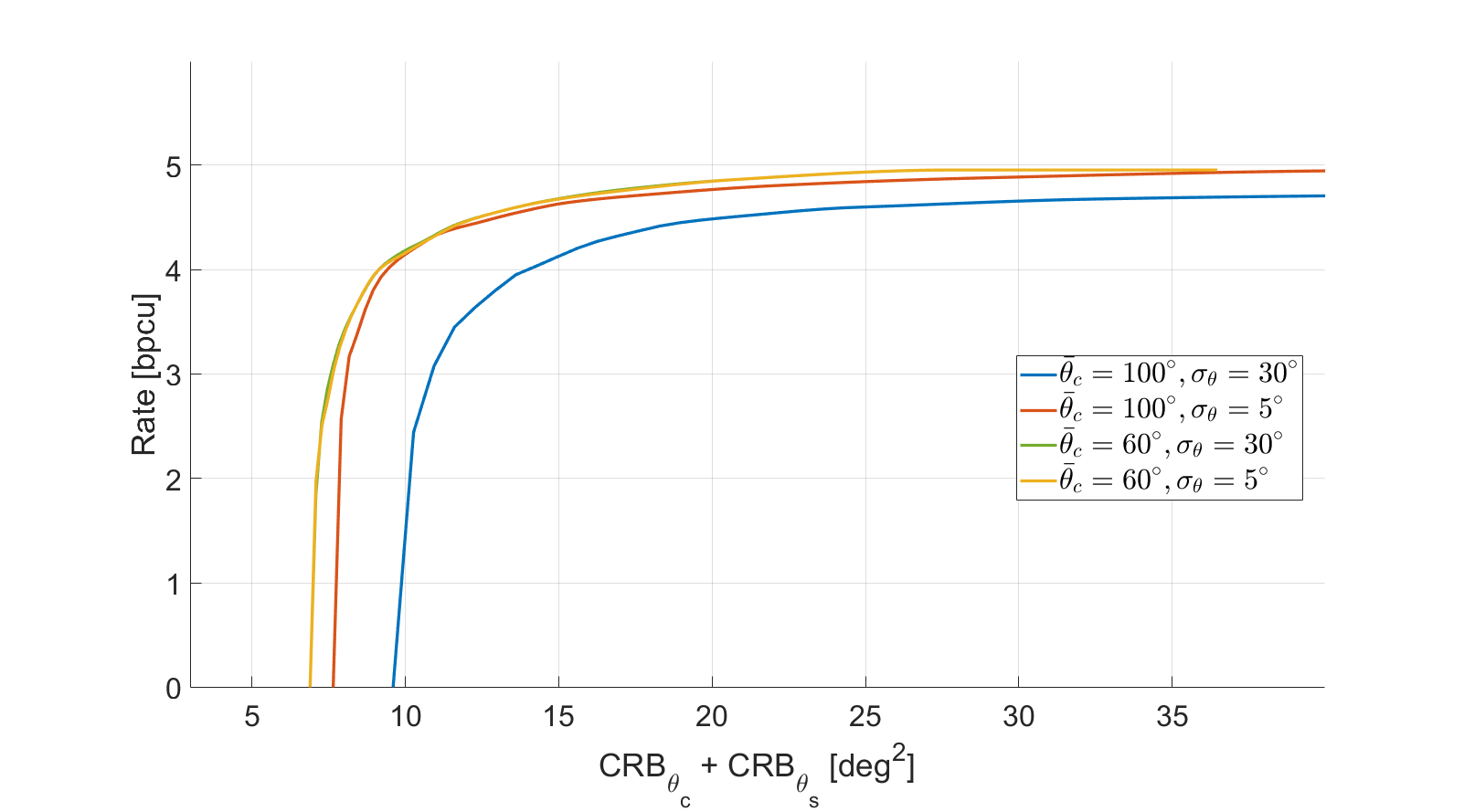}
    \caption{Comparison of achievable ISAC regions highlighting the impact of prior standard deviation $\sigma_{\theta}$ and angular separation between sensing and communication targets, with $\bar{\theta}_s$ fixed at $30^\circ$.
}
    \label{fig:comparison_ant_var}
\end{figure}

Fig.~\ref{fig:comparison_ant_var} shows the influence of key system parameters, specifically pre-acquisition variance and angular separation between targets, on the ISAC tradeoff. Regarding angular separation, for a fixed $\bar\theta_s =30^\circ$, the results demonstrate that smaller separations between targets ($\bar\theta_c = 60^\circ$) yield improved BCRB performance compared to widely separated targets ($\bar\theta_c = 100^\circ$). This finding suggests that the system's estimation capabilities are enhanced when targets are in closer proximity, potentially due to more direct beamforming advantages. For targets with greater angular separation ($\bar\theta_s = 30^\circ$, $\bar\theta_c = 100^\circ$), higher pre-acquisition variance significantly degrades sensing, as evidenced by the rightward shift of the blue curve. However, when targets are more closely spaced ($\bar{\theta}_s = 30^\circ$, $\bar{\theta}_c = 60^\circ$), the impact of variance becomes less pronounced, with the performance boundaries nearly coinciding regardless of the pre-acquisition variance.

\section{Conclusions} 
\label{sec:Conclusion}
In this paper, we analyzed the ISAC tradeoff in a practical scenario with multiple targets, where one target serves is a communication receiver. Under specific assumptions, we showed that the BCRB reduces to a sum of single-target estimation bounds and explored the optimal sensing and communication directions separately using convex optimization. Through various transmission strategies, we explored achievable BCRB-rate regions and identified effective strategies through power allocation. Future work could explore numerical generalization to higher number of targets as well as considering a broadcast scenario more communication receivers.

\section*{Acknowledgment}
This work has been supported in part by the National Science Foundation (NSF) under Grants No.~1900911 and No.~2312229.

\bibliographystyle{IEEEtran} 
\bibliography{refs} 

\end{document}